\documentstyle{article}
\input epsf

\begin{document}

\begin{flushright}
                    KEK CP-035~~~~~~~~\\
           KEK Preprint 95-126~~~~~~~~\\
\end{flushright}
\renewcommand{\baselinestretch}{1.5}
\large

\begin{center}
{\Large {\bf 
Test of QEDPS: A Monte Carlo for the hard photon\\
              distributions in $e^+e^-$ annihilation process}}
\vskip 0.5in
Y. Kurihara, J. Fujimoto, T. Munehisa(*) and Y. Shimizu
\vskip 0.2in

National Laboratory for High Energy Physics(KEK)
\vskip 0.05in
Oho 1-1 Tsukuba, Ibaraki 305, Japan

\vskip 0.2in
(*)Faculty of Engineering, Yamanashi University
\vskip 0.05in
Takeda, Kofu, Yamanashi 400, Japan

\end{center}
\vskip 0.5in
\begin{center}
{\bf ABSTRACT}
\end{center}

The validity of a photon shower generator QEDPS has been examined 
in detail. This is formulated based on the leading-logarithmic 
renormalization equation for the electron structure function and it 
provides a photon shower along the initial $e^{\pm}$. The main interest
in the present work is to test the reliability of the generator to
describe a process accompanying hard photons which are detected. 
For this purpose, by taking the $HZ^0$ production as the basic reaction,
the total cross section and some distributions of the hard photons are
compared between two cases that these photons come from either those 
generated by QEDPS or the hard process 
$e^+e^- \to HZ^0\gamma\cdots\gamma$. The comparison performed 
for the single and the double hard photon has shown a satisfactory 
agreement which demonstrated that the model is self-consistent. 
  
\eject

\def\GeV{\hbox{GeV}}

\noindent
\section{Introduction}

In a series of works\cite{ps1}-\cite{ps3} we have proposed a generator 
QEDPS which develops a photon shower along the initial $e^\pm$ of any 
hard annihilation process. This gives the correction by the initial state 
radiation(ISR). 
Usually ISR is implemented by the structure 
function of the initial $e^{\pm}$. 
By using the generator, however, it is possible not only
to give the corrected total cross section with arbitrary cuts but also
a distribution over any kinematical variable including transverse
momentum of photons. 
It should be also noticed that the 
annihilating $e^+$ and $e^-$ no more make the head-on collision, as 
they deviate from the beam axis by the radiation.

In the following sections we examine in detail the validity of
QEDPS. We take $e^+e^-\to HZ^0$ together with possible hard photons 
as the hard process. The next section
will make a brief review of QEDPS. We also summarize the analytic 
form of the structure function up to 
$O(\alpha^2)$, which should be compared against the photon shower in
section 3. 
Section 4 is devoted to the numerical test of QEDPS.
The correction to the total cross section is tested in the sub 
section 4.1 between the simulation and the analytic structure function. 
The subsequent two sub sections, 4.2 and 4.3, are 
devoted to a single and double hard photon emission, respectively. 
The last section gives the summary and discussions.

\section{QEDPS}
We review briefly the formulation of QEDPS in this section.
The basic assumption is that the structure function of an electron, 
with the virtuality $Q^2$ and the momentum fraction $x$, obeys the 
Altarelli-Parisi equation
\begin{equation}
{d D(x,Q^2)\over d\ln Q^2}={\alpha\over 2\pi}
                               \int_x^1 {dy \over y} P_+(x/y) D(y,Q^2),
                                        \label{eq:AP}
\end{equation}
in the leading-log(LL) approximation\cite{ll}. To solve this we modify 
the split-function as follows;
\begin{equation}
P_+(x)\simeq\theta(1-\epsilon-x)P(x)
             -\delta(1-x)\int\nolimits_0^{1-\epsilon}dyP(x),\quad
                             P(x)={ 1+x^2 \over 1-x }. 
\end{equation}
Here $\epsilon$ is a small quantity specified later. Then the original 
equation can be converted to the integral equation.
\begin{equation}
D(x,Q^2)= \Pi(Q^2,Q_s^2)D(x,Q_s^2)
+{\alpha\over2\pi}\int\nolimits_{Q_s^2}^{Q^2}{dK^2\over K^2}
    \Pi(Q^2,K^2)\int\nolimits_x^{1-\epsilon}{dy\over y}
           P(y)D(x/y,K^2).       \label{eq:intform}
\end{equation}
Here, rigorously speaking, $Q_s^2$ should be $m_e^2$ as it gives the 
initial condition. 
For simplicity the fine structure constant $\alpha$ is 
assumed not running with $Q^2$. The function $\Pi$, which is nothing 
but the Sudakov factor, is given by
\begin{equation}
 \Pi(Q^2,{Q'}^2) = \exp\left(- {\alpha\over 2 \pi} \int_{{Q'}^2}^{Q^2}
 { d K^2 \over K^2} \int_0^{1-\epsilon} d x P(x) \right),
  \label{eq:non}
\end{equation}
and implies the probability that the electron evolves from ${Q'}^2$ to 
$Q^2$ without emitting hard photon. In other words $\Pi$ already 
contains the contributions from the soft photon emission, which causes 
the change of the electron virtuality, and from the loop corrections 
in all orders of perturbation.

The integral form Eq.(\ref{eq:intform}) can be solved by iterating the 
right-hand side in a successive way. Then it is apparent that 
the emission of $n$ photons corresponds to the $n$-th iteration. Hence 
it is possible to regard the process as a stochastic one that suggests 
the following algorithm of the photon shower\cite{ps1}.

(a) Set  $x_b=1$. The variable $x_b$ is the fraction of the
    light-cone momentum of the virtual electron that annihilates.

(b) Choose a random number $\eta$. If it is smaller than 
    $\Pi(Q^2,Q_s^2)$, then the evolution stops. If not, one finds the 
    virtuality $K^2$ that satisfies $\eta =\Pi(K^2,Q_s^2)$ with which 
    a branching takes place.

(c) Fix $x$ according to the probability $P(x)$ between 0 and 
    $1-\epsilon$. Then $x_b$ is replaced by $x_b x$.
    One should go to (b) by substituting $K^2$ into $Q_s^2$ and 
    repeat until it stops.

Once an exclusive process is fixed by this algorithm, each branching
of a photon in the process is dealt with as a true process, that is, 
an electron with $x,K^2$ decays as
\begin{equation} 
      e^-(x,-K^2)\to e^-(xy,-{K'}^2)+\gamma(x(1-y),Q_0^2).
 \label{eq:elsplit}
\end{equation}
In the infinite momentum frame where the parent has the momentum 
$p=(E,{\bf 0}_T,p_z),$ $E=\sqrt{p_z^2-K^2},p_z=xp^*$, the daughter 
electron and the photon have $p'=(E',{\bf k}_T,yp_z)$ and 
$k=(k_0,-{\bf k}_T,(1-y)p_z)$, respectively, with 
$E'=\sqrt{y^2p_z^2+{\bf k}_T^2-{K'}^2}$ and 
$k_0=\sqrt{(1-y)^2p_z^2+{\bf k}_T^2+Q_0^2}$. Here we have introduced a 
cutoff $Q_0^2$ to avoid the infrared divergence. The momentum
conservation at the branching gives
\begin{equation}
     -K^2=-{K'}^2/y+Q_0^2/(1-y)+{\bf k}_T^2/(y(1-y)),
\end{equation}
which in turn determines ${\bf k}_T^2$ from $y,K^2,{K'}^2$. Hence one
can get the information on the transverse momentum. This enables us
to give the ${\bf k}_T^2$ distribution by the simulation as well as
the shape of $x$.

Further the kinematical boundary $y(K^2+Q_0^2/(1-y))\le{K'}^2$,
equivalent to ${\bf k}_T^2>0$, fixes $\epsilon$ as
\begin{equation}
         \epsilon=Q_0^2/{K'}^2,
\end{equation}
since $K^2\ll {K'}^2$ is expected.
In ref.\cite{ps3} it has been demonstrated that the existence of 
$\epsilon$ which depends on the virtuality gives rise to a small but 
non-negligible contribution to the structure function, though it is 
a constant. This is because when the virtuality becomes to the order of 
$Q_0^2$, $\epsilon$ is not small, while in Eq.(\ref{eq:intform}) it is 
assumed to be small enough.

The above description of the algorithm has shown the single cascade 
scheme.
This implies that either of $e^-$ or $e^+$ is able to radiate photons 
when the axial gauge vector is chosen along the momentum of the other 
electron, namely $e^+$ or $e^-$. In writing a computer code for the 
shower, however, it is convenient to use the double cascade scheme to 
ensure the symmetry of the radiation between $e^+$ and 
$e^-$\cite{double}. It can be shown that these two are mathematically
equivalent in the LL approximation. The extra finite contribution due 
to $\epsilon$ is, however, different between single and double cascade 
scheme. The details will be found in ref.\cite{ps3}.

In the formulation there are two parameters $Q_s^2$ and $Q_0^2$. 
In the program the following values are chosen:
\begin{equation}
    Q_s^2=m_e^2e=m_e^2\times2.71828\cdots,\qquad 
    Q_0^2=10^{-12}~~\hbox{GeV}^2.
\end{equation}
The former value was settled to take into account effectively 
the constant term $-1$ of $\beta$ in such a way
$\beta=(2\alpha/\pi)(\ln(s/m_e^2)-1)=(2\alpha/\pi)\ln(s/(m_e^2e))$.
Since the second parameter is unphysical, any physical observable 
should not depend on it. It has been checked that increasing
$Q_0^2$ up to $O(m_e^2/10$) leaves the result unchanged in the
statistical error of the event generation.

\vskip 0.5 in
\noindent
\section{Structure function}
  
The differential equation Eq.(\ref{eq:AP}) can be easily solved if one 
introduces the moment of the structure function by\cite{suppl}
\begin{equation}
D(n,s)=\int\nolimits_0^1dxx^{n-1}D(x,s).
\end{equation}
The solution is then
\begin{equation}
D(n,s)=\exp\left[{\beta\over2}\left({3\over2}-2{\rm S}_1(n-1)
                 -{1\over n}-{1\over n+1}\right)\right],
                             \label{eq:mom}
\end{equation}
with the initial condition $D(x,Q_s^2)=\delta(1-x)$. Here
\begin{equation}
{\rm S}_1(n)=\sum_{j=1}^\infty{n\over j(j+n)}.
\end{equation}
is an analytic continuation of the finite sum $\sum_{j=1}^n1/j$ into 
the complex $n$-plane. We expand Eq.(\ref{eq:mom}) with respect to
$\beta$ in the following way
\begin{eqnarray}
D(n,s)&\simeq&\left(1+{3\over8}\beta+{9\over128}\beta^2\right)
\exp\left(-{\beta\over2}{\rm S}_1(n-1)\right)     \nonumber\\
&&-{\beta\over4}\left({1\over n}+{1\over n+1}\right)
 \left(1-{\beta\over2}{\rm S}_1(n-1)\right)
-\left({\beta\over4}\right)^2{3\over2}
\left({1\over n}+{1\over n+1}\right)   \nonumber\\
&&+\left({\beta\over4}\right)^2{1\over2}
\left({1\over n}+{1\over n+1}\right)^2,
\end{eqnarray}
and further for $n\to\infty$
\begin{eqnarray}
\exp\left(-{\beta\over2}{\rm S}_1(n-1)\right)&\simeq&
\exp\left(-{\beta\over2}(\gamma_E+\ln n)\right)
-{\beta\over2}[{\rm S}_1(n-1)-\gamma_E-\ln n]\nonumber\\
&& +{1\over2!}\left({\beta\over2}\right)^2\{[{\rm S}_1(n-1)]^2
            -(\gamma_E+\ln n)^2\},
\end{eqnarray}
where $\gamma_E=0.57721\cdots$ is the Euler's constant.
The inverse transformation defined by
\begin{equation}
D(x,s)=\int\nolimits_{c-i\infty}^{c+i\infty}
               {dn\over2\pi i}x^{-n}D(n,s), \label{eq:strfxn}
\end{equation}
with $c>0$ being a real number to fix the integration path, gives
\begin{eqnarray}
  D(x,s)&=& \left[1+{3\over8}\beta
 +\left({9\over128}-{\zeta(2)\over8}\right)\beta^2\right]
                 {\beta\over2}(1-x)^{\beta/2-1} \nonumber \\
    && -{\beta\over 4} \bigl( 1+x \bigr) 
        -{\beta^2 \over 32} \biggl[ 4(1+x)\ln(1-x) 
              +{1+3x^2 \over 1-x}\ln x+(5+x) \biggr], 
                    \label{eq:Dxs}
\end{eqnarray}
Here we have made a resummation to get the first term 
$(1-x)^{\beta/2-1}$.

Eq.(\ref{eq:Dxs}) is used throughout for the test of QEDPS in
the following sections.
One should keep in mind that these 
results are obtained in the leading-log(LL) approximation and 
non-leading terms are not fully included corresponding to QEDPS.

\vskip 0.5in
\noindent
\section{Numerical test of QEDPS}
\subsection{Corrected total cross section}

In Ref.\cite{ps1} we have already compared the model with the 
analytic structure function\cite{rad} and also with the perturbative
calculations of the initial state radiation(ISR) available up to 
$O(\alpha^2)$\cite{suppl}. In these comparisons we had obtained a 
satisfactory agreement, but as explained in section 1
the recovery of the missing finite term should give better agreement. 
This is important to be confirmed since the experimental accuracy is 
becoming less than 1\%.

The basic hard process is 
\begin{equation}
e^+e^-\to HZ^0.     \label{eq:hz}
\end{equation}
We take this reaction as our example since it is not only the most 
suitable one to test the ISR corrections as there is no final state 
radiation but also very simple process with a few Feynman diagrams 
even when some extra photons are attached. Its cross section 
$\sigma_0(s)$ is combined with QEDPS to generate the events with 
indefinite number of photons. None of them is assumed to be observed, 
that is, no cut is imposed on the photons. On the other hand the 
corrected total cross section can be easily estimated by convoluting 
$\sigma_0(s)$ with the structure function. In this work all the hard 
process is calculated by Monte Carlo integration\cite{bases}. 
In other words we also have a generator for the hard process.

In Table 1 we show the results evaluated by these two different 
methods. The used parameters are given by
\begin{eqnarray}
&&   \alpha=1/137.036,~~ m_e=0.511\times10^{-3}~\mbox{GeV}, \nonumber\\
 &&   M_W=80.230~\mbox{GeV},~~ M_Z=91.188~\mbox{GeV}.  \label{eq:par}
\end{eqnarray}
The mass of Higgs particle is taken to be the same as $M_W$. This has 
no realistic meaning. The total energy is chosen in the range 
$\sqrt{s}=200-1000$ GeV, roughly from LEP200 to linear colliders.  
From the table one finds that the simulation agrees well with the 
analytic formula within the accuracy less than 0.4\%. This indicates that
QEDPS effectively reproduces the structure function which includes
the soft photon resummation and also the expansion of terms up to 
$O(\beta^2)$.

\begin{center}
\begin{tabular}{|c|c|c|c|}    \hline
$\sqrt{s}$(GeV) & $HZ$/ps & $HZ$/sf 
                & ($HZ$/ps)/($HZ$/sf)$-$1\\ \hline
200             & (5.727$\pm$0.004)$\times$10$^{-1}$ 
                & (5.727$\pm$0.002)$\times$10$^{-1}$ 
                &  0.0\%  \\ \hline
300             & (2.411$\pm$0.002)$\times$10$^{-1}$ 
                & (2.419$\pm$0.001)$\times$10$^{-1}$ 
                &  $-$0.33\% \\ \hline
400             & (1.193$\pm$0.002)$\times$10$^{-1}$ 
                & (1.196$\pm$0.001)$\times$10$^{-1}$ 
                &  $-$0.25\% \\ \hline
500             & (7.125$\pm$0.007)$\times$10$^{-2}$ 
                & (7.137$\pm$0.003)$\times$10$^{-2}$ 
                &  $-$0.17\% \\ \hline
\end{tabular}
\end{center}
\begin{center}
{\bf Table 1 } Total cross sections(pb) without cut. 
\end{center}

Finally we compare QEDPS with the fixed order perturbative calculations 
for the higher order corrections. At $\sqrt{s}=500$ GeV, for
example, the total cross sections are $\sigma_0=5.766\times10^{-2}$ pb 
in the tree level(Eq.(\ref{eq:hz})), $\sigma_1=7.181\times10^{-2}$ pb 
for the corrected one up to $O(\alpha)$ and 
$\sigma_2=7.208 \times10^{-2}$ pb 
up to $O(\alpha^2)$\cite{suppl,berends},
respectively. 
The numbers in Table 1 do not include the so called $K$-factor 
($K=1+\alpha/\pi(\pi^2/3 - 1/2)=1.00648 ...$) which recovers almost
the constant term in $O(\alpha)$. Multiplying this value, we get
$7.171 \times 10^{-2}$ pb of $HZ$/ps or
$7.183 \times 10^{-2}$ pb of $HZ$/sf, thus the convergence of the 
perturbation series shows a good behavior.
\vskip 0.5in
\noindent
\subsection{Single hard photon test}
   
Let us consider the following experimental situation. Together with
$H$ and $Z^0$ we shall observe a hard photon with $E_{\gamma}>1$ GeV  
outside of the cone of $5^\circ$ from the beam axis. Inside the cone 
any number of photons can be radiated. The photons with $E_{\gamma}<1$ 
GeV are regarded as soft and get no angular limitation. Hence one 
detects $HZ^0\gamma$ as the final state.
 
To estimate the total cross section and distributions with respect to
some kinematical variables, which we specify later, we consider four
different ways to estimate this cross section. The first one is simply 
to calculate it in the tree level
\begin{equation}
e^+e^-\to HZ^0\gamma.   \label{eq:hzg}
\end{equation}
Obviously this gives the lowest order result without any radiative 
correction. It is not expected to be precise enough then, but we
include this result in our comparison.

The second one is the same process but corrected up to $O(\alpha)$ 
in perturbation, which should give more accurate estimation. 
In this case one has to evaluate the cross section for the process
\begin{equation}
e^+e^-\to HZ^0\gamma\gamma,   \label{eq:hzgg}
\end{equation}
in the tree level under the experimental condition on the photons as
same as before. Also one needs the one-loop correction to the process 
Eq.(\ref{eq:hzg}). All the necessary calculations are found in 
Ref.\cite{suppl}. 

The third way is to impose the condition on the photons generated by
QEDPS being combined with $d\sigma_0(s)$. This is the same thing as given in
the previous section but with the cut. We denote this estimation as
$HZ/\gamma\hbox{ps}$ to imply that QEDPS is applied to the bare process
Eq.(\ref{eq:hz}) but one of the generated photons is observed.

The last one is to dress the photon shower to the radiative process 
Eq.(\ref{eq:hzg}) in the tree level. In this case the photon associated
with $HZ^0$ production is regarded as the observed one. Other photons
supplied by the simulation are supposed to be invisible so that they
escape inside of the cone or are soft. We denote this process as 
$HZ\gamma$/ps. Here it is important not to make
double counting of the hard photon. This can be achieved by the 
following manner. In Eq.(\ref{eq:elsplit}) the virtuality of the 
initial $e^{\pm}$ is required to satisfy $K^2 < {K'}^2$ at each branching in 
the cascade. This order should be also maintained for the electron in 
the process Eq.(\ref{eq:hzg}). In other words the virtuality of $e^\pm$ 
in the cascade is restricted to be smaller than that appears in the 
hard process.
  
One should notice that the comparison of the third and the fourth cases
provides a self-consistency check of the model while the first two 
allows the comparison with the perturbative calculations. 

The results of the total cross sections are summarized in Table 2 which
are obtained by these four different methods at various energies 
with the cut. 
\vspace {0.15cm}
\begin{center}
\begin{tabular}{|c|c|c|c|c|}    \hline
$\sqrt{s}$(GeV) 
                & $HZ\gamma$ (tree)
                & $HZ\gamma$ (up to $O(\alpha)$)
                & $HZ/\gamma$ps 
                & $HZ\gamma$/ps 
\\ \hline
200            
                &  5.602$\times$10$^{-2}$
                &  4.418$\times$10$^{-2}$ 
                & (4.534$\pm$0.008)$\times$10$^{-2}$ 
                & (4.424$\pm$0.002)$\times$10$^{-2}$ 
\\ \hline
300             
                & 3.284$\times$10$^{-2}$ 
                & 3.018$\times$10$^{-2}$ 
                & (3.086$\pm$0.016)$\times$10$^{-2}$ 
                & (2.945$\pm$0.002)$\times$10$^{-2}$ 
\\ \hline
400             
                & 1.765$\times$10$^{-2}$
                & 1.701$\times$10$^{-2}$
                & (1.754$\pm$0.010)$\times$10$^{-2}$ 
                & (1.648$\pm$0.001)$\times$10$^{-2}$ 
\\ \hline
500             
                &  1.101$\times$10$^{-2}$
                &  1.088$\times$10$^{-2}$
                & (1.143$\pm$0.007)$\times$10$^{-2}$ 
                & (1.048$\pm$0.001)$\times$10$^{-2}$ 
\\ \hline
\end{tabular}
\end{center}
\begin{center}
{\bf Table 2 } Total cross sections(pb) : single-hard photon required. 
\end{center}

At 500 GeV, the effect of $O(\alpha)$ corrections is small, less than 
1\%.
It does not mean, however, that Born approximation is enough
because the various distributions are different between them
as shown in Fig.1.  
\begin{figure*}[htb]
\centerline{\epsfbox{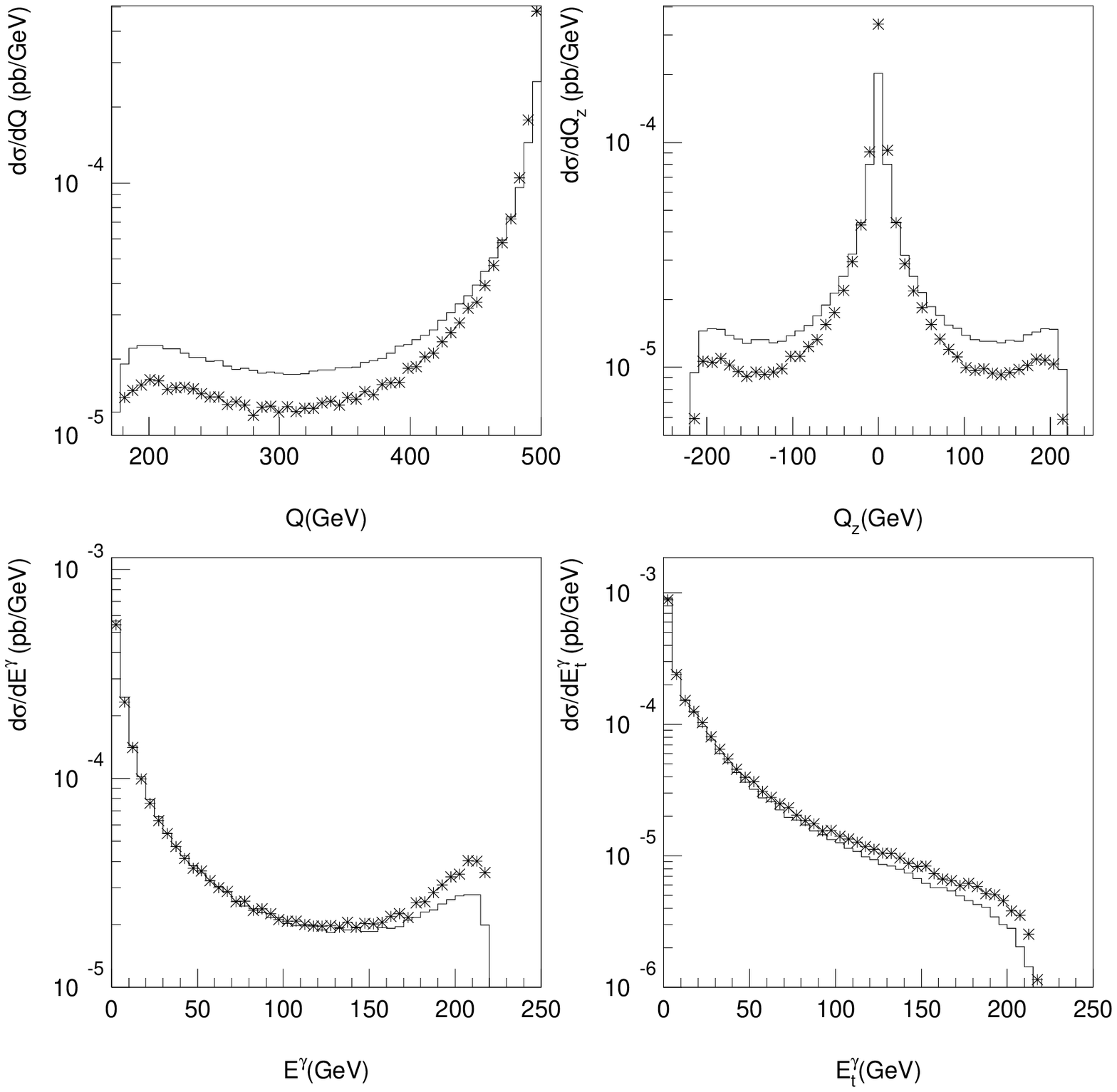}}
\caption{}
\end{figure*}
The plots show the first case and the histograms does the second one.
$Q$ is the virtuality of the $s$-channel virtual boson($Z^0$) or the
invariant mass of the $HZ^0$ system. $Q_z$  is the longitudinal momenta 
of this system.  $E^{\gamma}$ and $E_{t}^{\gamma}$ are the energy and 
the transverse momentum of the required hard photon with respect to the 
beam axis, respectively.
There is seen a large discrepancy in the $Q$ and $Q_z$ distributions.
This is not surprising because there is no correction from the soft 
photon emission in the tree level case. On the other hand, $E^{\gamma}$ 
and $E_t^{\gamma}$ distributions agrees rather well.
   
Next we show in Fig.2 the comparison of the same distributions between
the third case, $HZ/\gamma$ps and the second case, $HZ\gamma$ with 
the $O(\alpha)$ corrections.
Here, again, the histograms show the second case and the plots 
the third one.
\begin{figure*}[htb]
 \centerline{\epsfbox{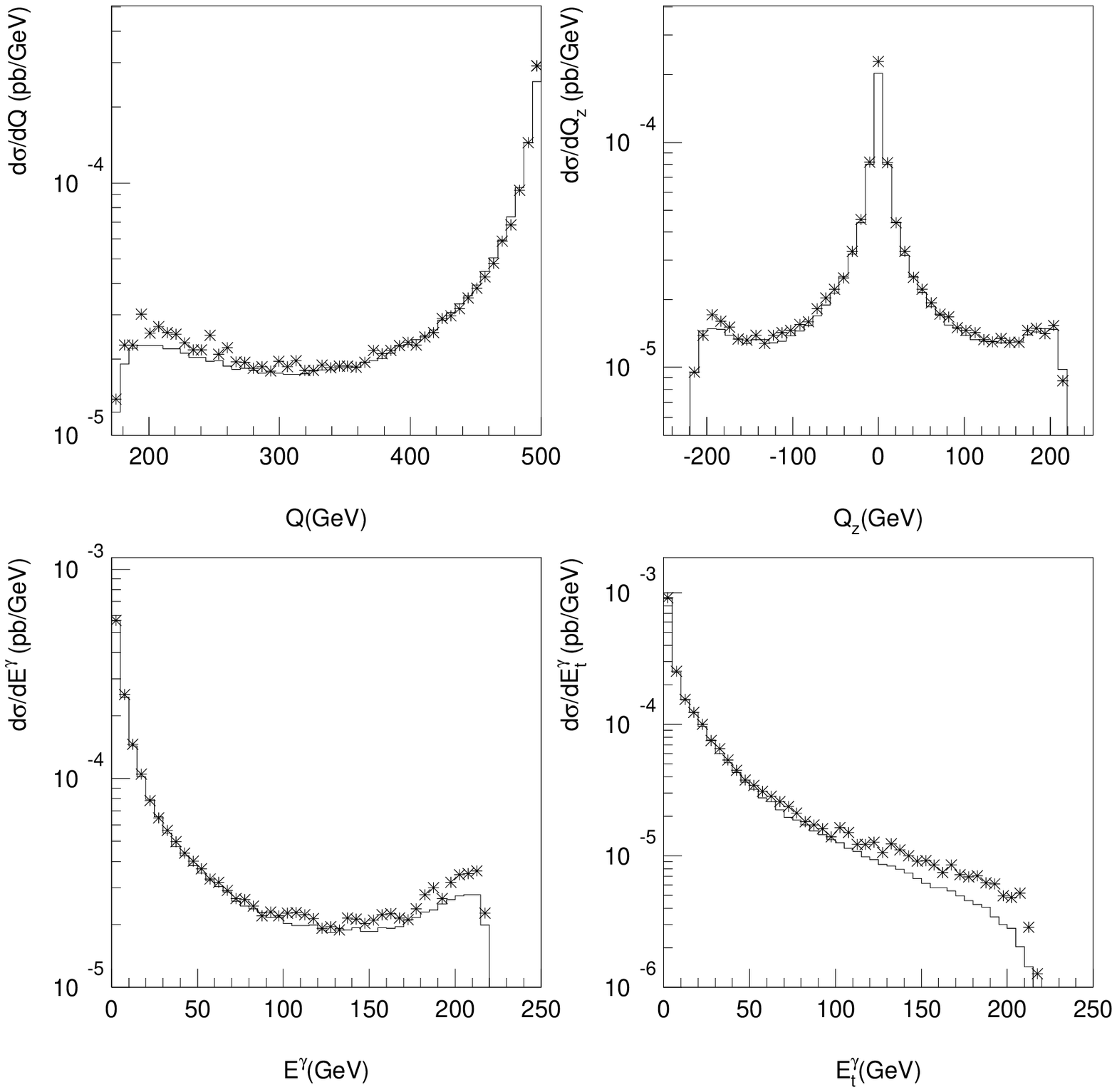}}
\caption{}
\end{figure*}
This time $Q$ and $Q_z$ distribution are quite consistent because 
the both cases include the soft photon 
emission. In $E^{\gamma}$ and $E_t^{\gamma}$ 
distributions the third case shows an enhancement against to the second,
but still consistent. One reason of this enhancement  is presumably
attributed to the fact that the constant terms of $O(\alpha/\pi)$ are
missing in the LL approximation.
Because of this enhancement, the total cross section
of $HZ/\gamma$ps are always larger than that for $HZ\gamma$ with the
$O(\alpha)$ corrections. QEDPS model, however, gives the events with 
the acurracy in the level of a few percent even though a hard photon 
is required. 

Fig.3 shows the same comparison as in Figs.1 and 2 but between
the fourth case, $HZ\gamma$/ps, given by the plots and 
the second, $HZ\gamma$ with $O(\alpha)$ corrections, by the histograms.
The agreement of all the distributions is very well.
But the total cross sections in Table 2 of the fourth case are
always smaller than the second one. This corresponds to the fact that
in the lowest edge of the $E^{\gamma}_t$, the differential cross 
section of the fourth case is smaller than the
second one. As mentioned above, we require the order of the virtuality
to combine the one photon emission process with the QEDPS model. 
This recipe may be still naive and has some ambiguity
in the very small $E^{\gamma}_t$ region. On the other hand, in the
large $E^{\gamma}_t$ region, it works well.
As long as  one intends to study the events with a hard
photon, the fourth is the best one. 

\begin{figure*}[htb]
\centerline{\epsfbox{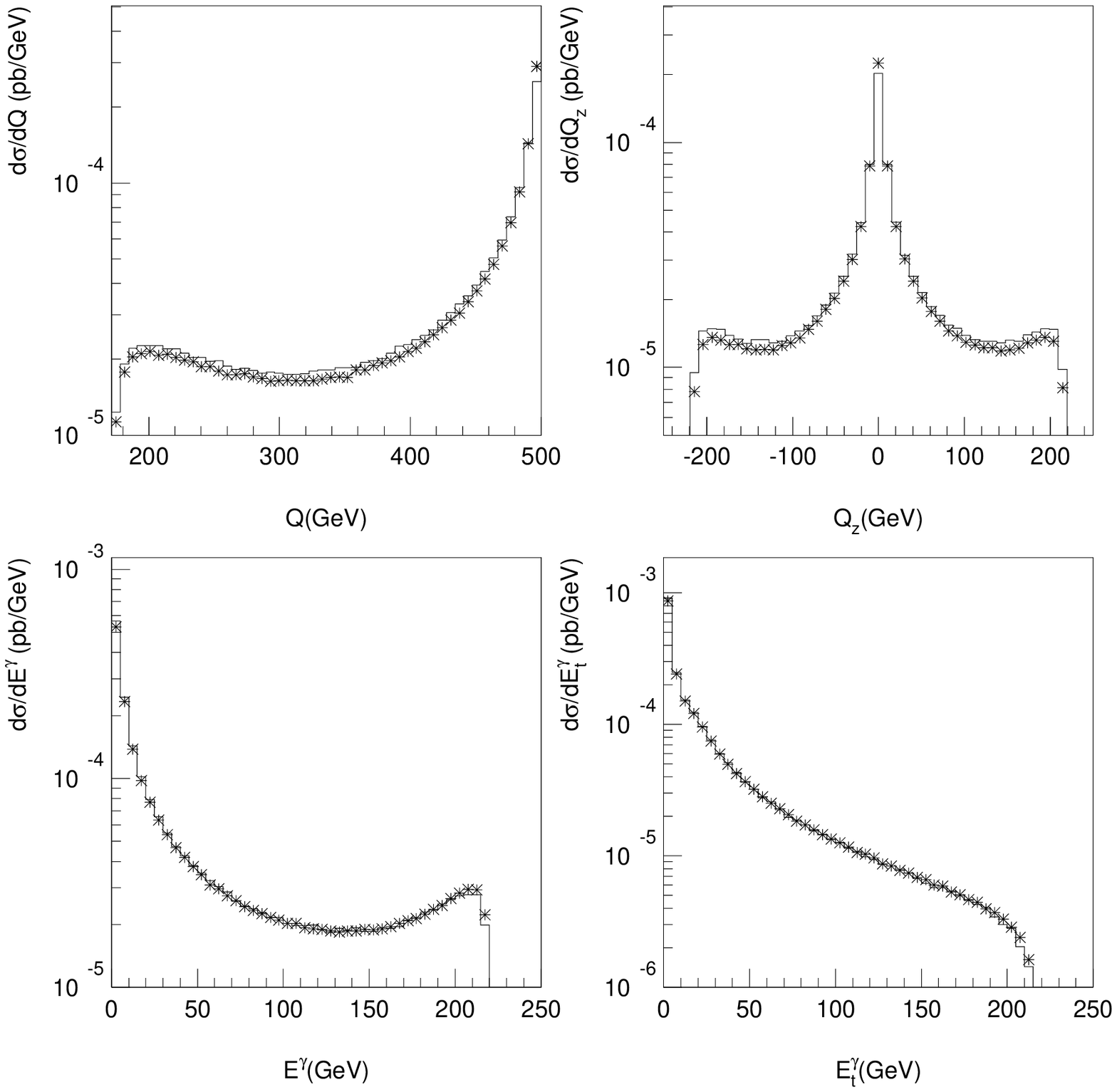}}
\caption{}
\end{figure*}
From these results one can conclude that QEDPS is consistent not only 
with the result of perturbation calculation up to the $O(\alpha)$ 
corrections but also within itself.  This is a strong support
for the reliability of QEDPS.

What happens if one convolutes the structure function with the radiative
process Eq.(\ref{eq:hzg}), that is, $HZ\gamma$/sf? It is obvious that
one has the double-counting because the hard photon is also contained 
in the structure function. Hence a naive convolution leads to 
an overestimation and one should introduce some special prescription 
to get a correct result\cite{olden}.

It would be, however, interesting to see if QEDPS could reproduce 
the naive $HZ\gamma$/sf. In the former, as explained in the previous 
section, the virtuality cut prevents the double-counting. Then one may
expect that to remove this cut would give the same result as 
$HZ\gamma$/sf. One kinematical difference should be noticed. In the case
of the structure function the head-on collision of $e^\pm$ takes place.
On the other hand, with QEDPS the annihilation of electrons in the hard 
process cannot make the head-on collision because an emitted photon
has a finite $p_t^\gamma$. To compare these two on the same ground then,
we require the hard photon cut in {\it the CM system of the annihilating} 
$e^{\pm}$. The results summarized in Table 3 show that the above 
consideration is legitimate. The total cross section in the first row 
calculated by the structure function does not show any significant 
difference between Lab. and CM system while that by QEDPS in the second
row considerably 
changes, resulting the consistent number with $HZ\gamma$/sf only where 
the cut is imposed in the CM system. These results are unphysical after 
all, as the double-counting is allowed and the cut in the CM system is 
artificial. The comparison, however, provides another evidence that 
QEDPS is consistent with the structure function.
    
\begin{center}
\begin{tabular}{|c|c|c|}    \hline
hard photon cut & in Lab. sys. & in CM sys. \\ \hline
 $\sigma(HZ\gamma$/sf) & (1.256$\pm$0.002)$\times$10$^{-2}$ 
              & (1.266$\pm$0.001)$\times$10$^{-2}$ \\ \hline
 $\sigma(HZ\gamma$/ps) without&  & \\
 the virtuality cut     
&\multicolumn{1}{c|}{\raisebox{1.5ex}[0pt]
                     {(1.415$\pm$0.005)$\times$10$^{-2}$}} 
&\multicolumn{1}{c|}{\raisebox{1.5ex}[0pt]
                     {(1.260$\pm$0.001)$\times$10$^{-2}$}} \\ \hline
\end{tabular}
\end{center}
\begin{center}
{\bf Table 3 } Total cross sections(pb) at $\sqrt{s}$=500GeV :
               single-hard photon required. 
\end{center}

\vskip 0.5in
\noindent
\subsection{Double hard photon test}

Further check of the model is possible by requiring the double hard 
photon emission. In this case, however, the perturbative calculation
is restricted to only the tree process, Eq.(\ref{eq:hzgg}), as we have
no loop-correction of $O(\alpha^3)$ yet. Hence the first estimation 
comes from this tree process.

The second one is to dress the photon shower to the process 
Eq.(\ref{eq:hz}) in the similar way as the third case in the 
previous section, It means that the two hard photons are generated
by QEDPS.

The third one is to apply QEDPS to the process Eq.(\ref{eq:hzgg}). 
In this case, two hard photons originate from the hard process 
Eq.(\ref{eq:hzgg}) and others from the simulation are not observed.

The cuts for the hard photon is the same as before. Table 4 shows that 
even though the double-photon is required, QEDPS works in a consistent 
way independent of the allocation of the hard photon.
\normalsize
\begin{center}
\begin{tabular}{|c|c|c|c|c|}    \hline
$\sqrt{s}$(GeV) & $HZ\gamma\gamma$ 
                & $HZ/\gamma\gamma$ps 
                & $HZ\gamma\gamma$/ps 
                \\ \hline
200             & (2.120$\pm$0.004)$\times$10$^{-3}$
                & (1.644$\pm$0.018)$\times$10$^{-3}$ 
                & (1.640$\pm$0.003)$\times$10$^{-3}$ 
\\ \hline
300             & (2.161$\pm$0.004)$\times$10$^{-3}$ 
                & (2.040$\pm$0.017)$\times$10$^{-3}$ 
                & (1.922$\pm$0.004)$\times$10$^{-3}$ 
\\ \hline
400             & (1.384$\pm$0.003)$\times$10$^{-3}$ 
                & (1.370$\pm$0.011)$\times$10$^{-3}$ 
                & (1.257$\pm$0.003)$\times$10$^{-3}$
\\ \hline
500             & (9.481$\pm$0.014)$\times$10$^{-4}$
                & (9.883$\pm$0.062)$\times$10$^{-4}$ 
                & (8.707$\pm$0.022)$\times$10$^{-4}$
\\ \hline
\end{tabular}
\end{center}
\large
\begin{center}
{\bf Table 4 } Total cross sections(pb) : double-hard photon required. 
\end{center}

Fig.4 shows the comparison in the distributions. In this
figure, the histograms are for $HZ/\gamma\gamma$(tree) and the cross and 
the circle correspond to $HZ/\gamma\gamma$ps and $HZ\gamma\gamma/$ps, 
respectively. 
There appears a discrepancy in the distributions of $Q$ and
$Q_z$ between the Born and the other two with QEDPS 
because the higher order corrections are missing in the first case.
On the other hand, in the region of high $Q$, 
$HZ/\gamma\gamma$ps and $HZ\gamma\gamma/$ps are quite consistent.
In the low region of $Q$, however, there seems a difference. 
Since there is no perturbative calculation with higher order we cannot 
say definitely which distribution is better but it is known in the study
of QCD parton shower that the LL approximation always gives bigger
estimation than the exact one at larger momentum transfer region. 
$E^{\gamma}$ and $E^{\gamma}_t$ distributions agree well among 
three cases. As a conclusion, these results show the
self-consistency of the QEDPS model again even for the double hard photons.
\begin{figure*}[htb]
\centerline{\epsfbox{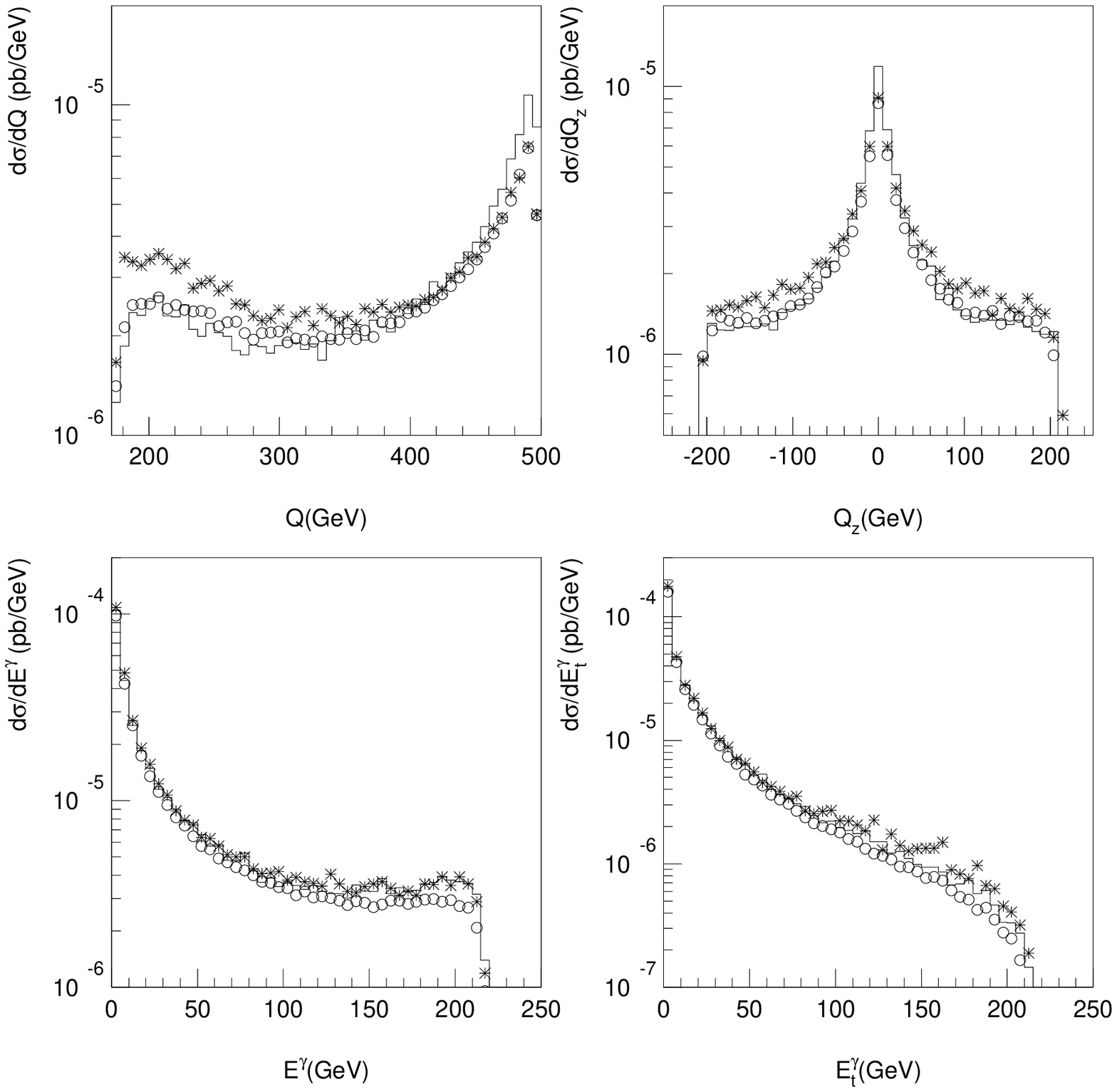}}
\caption{}
\end{figure*}
\vskip 0.5in
\noindent
\section{Summary and discussions}

In this work we have tested QEDPS, a generator for the radiative 
corrections in $e^+ e^- $ annihilation. The total cross section 
corrected by QEDPS agrees with that calculated by the analytic formula 
of the structure function with the accuracy less than around 0.3\%
in the energy range of LEP200 to linear colliders.
We checked the hard photon distribution of QEDPS against the calculation
of the matrix element with the $O(\alpha)$ corrections and got
an agreement. In order to make the self-consistency 
check, we applied QEDPS not only to $e^+e^- \rightarrow HZ^0$
but also to $e^+e^- \rightarrow HZ^0\gamma$ and 
$HZ^0\gamma\gamma$. In these cases we have introduced the virtuality 
cut to avoid the double-counting of the hard photon. As the result it
has been shown that QEDPS is self-consistent in various distributions. 
Through all of these studies we 
conclude that the model for the photon showers for the ISR is 
established fairly well even in the LL approximation.
This is also demonstrated for more complicated process such as
$e^+e^- \rightarrow \mu^- \nu_{\mu}u\bar{d}$ as described in 
Ref.\cite{lep200}.

In the present work we have restricted the radiation only to ISR. 
The extension to include the final state radiation(FSR) is 
straightforward. In this case, however, a problem emerges how to 
incorporate with the interference between ISR and FSR, which is, in
general, not so small to be ignored safely. Also the extension to the 
next-to-leading-logarithm(NLL) is urgent to make more precise 
predictions. This can be done in the same way as that in QCD parton 
shower\cite{nll}. These problems are now under investigation.

\eject
{\bf Acknowledgements}

We would like to thank our colleagues of KEK working 
group(Minami-Tateya) and those in LAPP for valuable discussions.
Particularly we are indebted to F. Boudjema, G. Coignet, T. Kaneko and D.
Perret-Gallix for their continuous interest and encouragement.
This work has been done under the collaboration between KEK and LAPP 
supported by Monbusho, Japan(No. 07044097) and CNRS/IN2P3, France.

\eject

\eject

{\bf Figure Captions}

Fig.1~~ Comparison of the distributions between $HZ\gamma$(tree) and
$HZ\gamma$ with up to the $O(\alpha)$ corrections  at 
$\sqrt{s}$=500GeV. The plots show the former
and the histograms the latter.
$Q$ and $Q_z$ are the virtuality and 
the longitudinal momenta of the $s$-channel
virtual boson, respectively. 
$E^{\gamma}$ and $E_{t}^{\gamma}$ are the energy and the transverse 
momentum of the required hard photon with respect to the beam axis, 
respectively.

Fig.2~~ Comparison of the distributions between $HZ\gamma$ 
with up to the
$O(\alpha)$ corrections(histograms) and
$HZ/\gamma$ps(plots) at $\sqrt{s}$=500GeV. 

Fig.3~~ Comparison of the distributions between $HZ\gamma$ 
with up to the
$O(\alpha)$ corrections(histograms) and
$HZ\gamma$/ps(plots) at $\sqrt{s}$=500GeV. 

Fig.4~~ Comparison of the distributions among 
$HZ\gamma\gamma$(histogram), $HZ/\gamma\gamma$ps (star plots) 
and $HZ\gamma\gamma$/ps (circle plots) at $\sqrt{s}$=500GeV. 
The variables $Q$, $Q_z$, $E^{\gamma}$ and $E_{t}^{\gamma}$ are the 
same as in Fig. 1.
 
\end{document}